# Longitudinal Data with Follow-up Truncated by Death: Match the Analysis Method to Research Aims

Brenda F. Kurland, Laura L. Johnson, Brian L. Egleston and Paula H. Diehr

*Abstract.* Diverse analysis approaches have been proposed to distinguish data missing due to death from nonresponse, and to summarize trajectories of longitudinal data truncated by death. We demonstrate how these analysis approaches arise from factorizations of the distribution of longitudinal data and survival information. Models are illustrated using cognitive functioning data for older adults. For *unconditional* models, deaths do not occur, deaths are independent of the longitudinal response, or the unconditional longitudinal response is averaged over the survival distribution. Unconditional models, such as random effects models fit to unbalanced data, may implicitly impute data beyond the time of death. *Fully conditional* models stratify the longitudinal response trajectory by time of death. Fully conditional models are effective for describing individual trajectories, in terms of either aging (age, or years from baseline) or dying (years from death). Causal models (principal stratification) as currently applied are fully conditional models, since group differences at one timepoint are described for a cohort that will survive past a later timepoint. *Partly conditional* models summarize the longitudinal response in the dynamic cohort of survivors. Partly conditional models are serial cross-sectional snapshots of the response, reflecting the average response in survivors at a given timepoint rather than individual trajectories. *Joint* models of survival and longitudinal response describe the evolving health status of the entire cohort. Researchers using longitudinal data should consider which method of accommodating deaths is consistent with research aims, and use analysis methods accordingly.

*Key words and phrases:* Censoring, generalized estimating equations, longitudinal data, missing data, quality of life, random effects models, truncation by death.

Brenda F. Kurland is Ph.D. Assistant Member, Clinical Research Division, Fred Hutchinson Cancer Research Center, 1100 Fairview Ave. N. D5-360, P.O. Box 19024, Seattle, Washington 98109-1024, USA e-mail: bkurland@fhcrc.org. Laura L. Johnson is Ph.D. Statistician, Office of Clinical and Regulatory Affairs, National Center for Complementary and Alternative Medicine, National Institutes of Health, 6707 Democracy Blvd. Suite 401, MSC 5475 Bethesda, Maryland 20892-5475, USA, e-mail: johnslau@mail.nih.gov. Brian L. Egleston is Ph.D. Associate Member, Biostatistics Facility, Fox Chase Cancer Center, 333 Cottman Ave., Philadelphia, Pennsylvania 19111-2497, USA e-mail: Brian.Egleston@fccc.edu. Paula H. Diehr is Ph.D. Professor, Departments of Biostatistics and Health Services, Box 357232 University of Washington, Seattle, Washington 98195-7232, USA e-mail: pdiehr@u.washington.edu.

## 1. INTRODUCTION

Research studies often collect information at multiple timepoints. For example, the Cardiovascular Health Study (CHS), an observational study of 5888 older adults, conducted annual assessments of cardiovascular functioning and other health measures for up to 18 years. With these longitudinal measurements, CHS data have been used to study disease course (Kaplan et al., 2005), health in the years leading up to a diagnosis (Diehr et al., 2001b) and the natural history of aging (Burke et al., 2001; Diehr et al., 2002).

Missing data can be an impediment to interpreting longitudinal data. Suppose a cohort of 200 subjects report self-rated health at age 70 years, and only 150 of these subjects are located for follow-up. If the average self-rated health at age 75 is higher than the average at 70, the increase could reflect improvement in individuals' health, attrition of sicker participants or death of sicker participants.

Many analysis methods for longitudinal data with dropout and nonresponse have been proposed to address different research aims, study designs, missing data patterns and estimation strategies (Little and Rubin, 1987; Rubin, 1987; Robins, Rotnitzky and Zhao, 1995). Comparatively little work in statistical methodology has addressed data missing when deaths occur during the period of follow-up, and recent work has focused mostly on principal stratification (Frangakis and Rubin, 2002; Frangakis et al., 2007). Kurland and Heagerty (2005) characterized targets of inference for longitudinal data truncated by death by factorizations of the joint distribution of survival and longitudinal response, $f(S, Y)$. The factorization was introduced primarily to provide context for a single approach, the partly conditional mean model. In this article, we explore several targets of inference in detail, and give guidance on appropriate analysis techniques for common scientific questions arising for longitudinal data truncated by death. We present six modeling options, illustrated using both hypothetical and actual CHS data. The hypothetical data without measurement error illustrate clearly how modeling choices for longitudinal data reflect assumptions about survival. The real data illustrate how standard analysis techniques such as random effects models and generalized estimating equations (GEE) may be applied to address different research aims involving longitudinal data truncated by death. Bias, estimation and efficiency are important issues for data analysis. However, we focus only on our primary interest, the *interpretation* of regression model estimands. Although each model is used to fit a slope and expected response values, these estimands for the longitudinal response are "apples and oranges," not directly comparable due to different factorizations of longitudinal response and survival.

## 2. BACKGROUND: FOLLOW-UP CENSORED BY NONRESPONSE (DROPOUT)

A brief review of models for longitudinal data with monotone dropout provides a foundation for discussing analysis of longitudinal data truncated by death. Two common, widely applied analysis methods for longitudinal data are random effects models (Laird and Ware, 1982) and generalized estimating equations (GEE) (Liang and Zeger, 1986). By modeling a structure for the correlation between subjects' longitudinal responses, a correctly specified random effects model will yield consistent, unbiased estimates of regression parameters by maximum likelihood estimation, even with unbalanced data (Laird, 1988). For example, if sicker participants drop out, their trajectory of decline in self-rated health is continued implicitly by a well-specified random effects model. If trends for dropouts can be inferred from observed data and parameters for longitudinal response and dropout are distinct (missing at random, such as when scores decline before dropout), the missingness is ignorable and the overall rate of change may be analyzed as if no one has dropped out. If the decline in health that leads to dropout starts after the last recorded measurement, then dropout is nonignorable, and random effects models are not an easy solution. Untestable assumptions must be made about nonignorable dropout processes to model longitudinal trends (Laird, 1988). GEE can accommodate data missing at random if estimating equations are weighted by the inverse probability of dropout (Robins, Rotnitzky and Zhao, 1995). Giving additional weight to observed data for people who were likely to drop out is similar to implicit or explicit imputation of unobserved data.

In fact, under some conditions, weighted GEE and imputation will give the same results (Paik, 1997). Missing at random is often a reasonable assumption, especially when longitudinal observations are closely spaced relative to mechanisms acting on both dropout and response. For example, preclinical cognitive changes could likely be detected by annual assessments before a CHS participant becomes impaired by dementia in a way that would lead to nonresponse. However, analysis of longitudinal data with MAR dropout still requires accurate modeling of the regression model (fixed effects) and either correlation (for random effects models) or dropout (for weighted GEE) (Kurland and Heagerty, 2004).

## 3. DATA EXAMPLES AND NOTATION

### 3.1 Cardiovascular Health Study (CHS)

The Cardiovascular Health Study (CHS) was a population-based prospective longitudinal study of 5888 adults aged 65 years and older at baseline (Fried et al., 1991). Cognitive functioning was assessed annually for up to 10 years by the Modified Mini-Mental State Examination (3MSE, scored from 0 to 100) (Teng and Chui, 1987). Our primary goal in the CHS analysis is to describe the trajectory of cognitive functioning (3MSE) over time, and to estimate 3MSE scores at different ages. Gender effects are explored in each analysis to add a between-person variable of interest to the longitudinal (within-person changes) model, and to explore the effects of differential survival on different regression approaches for longitudinal and survival data. We examine the 3814 participants aged 70 years and older at baseline. In this cohort, 1356 participants (36%) died during follow-up: 44% (744/1703) of men and 29% (612/2111) of women. We will examine how different analysis methods each yield 3MSE trajectories (rates of change) and fitted values (expected cognitive status at specific ages), but address different research aims.

Although models may be constructed to accommodate both deaths and nonresponse (Kurland and Heagerty, 2005), we imputed data missing due to nonresponse for simplicity of presentation. Data were not considered missing if follow-up was truncated by death, or censored by the end of the study period. (A later cohort to boost minority recruitment received 6 annual assessments instead of 10.) Most participants ($n = 2061$, 54%) completed all scheduled 3MSE assessments. About 17% of 3MSE scores (5174 of 31093) were missing due to participant nonresponse, but most participants with missing data had only one or two scores missing ($n = 948$ participants). Some participants had dropped out of the study, and were missing 7 or more scores each ($n = 134$). Nonresponse was accommodated by single imputation using the Markov Chain Monte Carlo method of PROC MI in the SAS/STAT software, version 9.1 (SAS Institute, Inc., Cary, NC). Imputation was stratified by time of death (for decedents) and recruitment group (for nondecedents), and was modeled based on observed 3MSE values, baseline age, gender and recruitment group.

TABLE 1
*Longitudinal 3MSE scores for 4 hypothetical CHS participants (X = deceased)*

| Participant | Age | | | | | |
|---|---|---|---|---|---|---|
| | 70 | 71 | 72 | 73 | 74 | 75 |
| A ("normal") | 90 | 90 | 90 | 90 | 90 | 90 |
| B ("mild cognitive impairment") | 84 | 82 | 80 | 78 | 76 | 74 |
| C ("terminal decline") | 84 | 80 | 76 | X | X | X |
| D ("terminal decline") | 65 | 50 | 35 | X | X | X |

Accommodation of deaths in CHS data also will be described using simplified hypothetical data. Table 1 shows 3MSE data for 4 hypothetical participants with a baseline age of 70. Participant A, representing normal cognitive functioning, has a 3MSE of 90 points at all assessments. Participant B's linear decline from 84 to 74 points over 5 years reflects a possible trajectory of mild cognitive impairment (which could be interpreted as preclinical Alzheimer's disease). Participants C and D both decline between baseline and age 72, and die before age 73.

### 3.2 Notation

Vector $Y_i$ represents the longitudinal response (i.e., cognitive functioning or quality of life), measured at multiple timepoints for participant $i$. The dimension of $Y_i$ may differ for individuals (values of $i$), due to death. For example, in Table 1 (hypothetical CHS data), responses for participant A, $Y_A$, are the vector $(90, 90, 90, 90, 90, 90)$, and for participant C, $Y_C$, are $(84, 80, 76)$. Scalar variable $S_i$ represents survival time for participant $i$, such as age at death or weeks from baseline until death: in Table 1, $S_C$ is 73 years. The dimension of $Y_i$ is determined by the value of $S_i$, but is not in principle affected by data missing due to nonresponse. In practice, the analysis in Section 4 uses imputation of data missing due to nonresponse to ensure that the observed response vector is of the proper dimension. The joint distribution $f(Y_i, S_i)$ describes the probability that $Y_i$ takes on a vector of specific values, and that participant $i$ dies at a specific time. We assume that $(Y_i, S_i)$ are independent and identically distributed over $i$.

TABLE 2
*Summary of statistical models for longitudinal response and survival (time of death)*

| | Statistical model | Sample research setting | Primary analysis method | Comments |
|---|---|---|---|---|
| a. | *Unconditional* $f(Y_i)$ Describe $Y_i$ (longitudinal response) in setting where survival ($S_i$) is unrelated to $Y_i$, or when death does not result in missing data | Rate of local recurrence following ablation of liver tumors | Mixed effects / random effects / latent variable regression | May implicitly impute data beyond death |
| b. | *Fully conditional: pattern-mixture* $f(Y_i \mid S_i = s)$ Describe $Y_i$ separately for groups defined by survival time | Longitudinal change in physical functioning following stroke, separately for $< 1$, 1–5, and $> 5$-year survivors | Mixed effects / random effects regression stratified by survival time | Describes individual trajectories, but uses future survival information to predict earlier responses |
| c. | *Fully conditional: principal stratification* $f(Y_i(z) \mid S_i(0) > s, S_i(1) > s)$ Describe average causal treatment effect for stratum that would survive past time $s$ regardless of treatment | Average QOL difference for toxic treatment with greater survival, versus less toxic treatment with lesser survival, in subjects expected to live 6+ months on either treatment | Weighted averaging of estimated outcomes from models such as generalized linear models | Weights are unidentifiable and must be explored through strong assumptions and/or sensitivity analysis |
| d. | *Fully conditional: terminal decline* $f(Y_i \mid S_i = s)$ Describe $Y_i$ counting backward from time of death | Terminal decline studies | Mixed effects / random effects regression | Time scale is retrospective |
| e. | *Partly conditional: regression conditioning on being alive* $f(Y_i \mid S_i > t)$ Describe $Y_i$ in the dynamic cohort of survivors at each timepoint | Average physical functioning in survivors at 6 months and 5 years after stroke | GEE with independence working correlation | Describes longitudinal trend of dynamic cohort, not individuals |
| f. | *Joint model* $f(Y_i, S_i)$ Describe both $Y_i$ and $S_i$ for example, "probability of being alive and healthy" | Percent of stroke patients who are alive and can perform self-care 6 months after stroke | Logistic regression, GEE (binary outcome); specialized multiple response methods | Continuous longitudinal outcomes may need to be categorized for analysis |

## 4. STATISTICAL MODELS FOR LONGITUDINAL RESPONSE AND SURVIVAL

Regression models for longitudinal data describe the relationship between predictors and the longitudinal response, $Y_i$. Because survival $S_i$ determines the length of $Y_i$ and is not fixed, regression models of longitudinal data truncated by death must explicitly or implicitly model survival as well. A single regression model could be built for the joint distribution $f(Y_i, S_i)$, or for factorizations based on the definitions of joint and conditional distributions: $f(Y_i \mid S_i) f(S_i)$ or $f(S_i \mid Y_i) f(Y_i)$. We will characterize models for $Y_i$ as unconditional, fully conditional or partly conditional based on how, or whether, the longitudinal response model conditions on $S_i$. These models are defined in more detail below, and summarized in Table 2. Each model is applied to the hypothetical (Table 3) and actual (Figure 1) CHS data. Models are fit to data for all participants, but

TABLE 3
*Hypothetical CHS data (Table 1) estimated age 75 3MSE score and 3MSE slope, using models that accommodate deaths in different ways. Even a predicted value for the response, such as "3MSE at age 75," represents a different estimand for each model*

| | Sample research question(s) | 3MSE at age 75 | Linear 3MSE slope (annual change in 3MSE) |
|---|---|---|---|
| a. | *Unconditional* $f(Y_i)$<br>What is the expected 3MSE at age 75, in an immortal cohort? | $\frac{90+74+64+(-10)}{4} = 54.5$ points | $\frac{0+(-2)+(-4)+(-15)}{4} = -5.25$ points per year |
| b. | *Fully conditional: survivors* $f(Y_i\|S_i > 75)$<br>What is the expected 3MSE at age 70 or at age 75, for people who live to be at least 75?<br>*Fully conditional: decedents* $f(Y_i\|S_i \leq 75)$<br>What is the expected 3MSE at age 70 for people who die at age 71–75? | $\frac{90+74}{2} = 82$ points<br><br><br>(both deceased at age 75) | $\frac{0+(-2)}{2} = -1.0$ points per year<br><br><br>$\frac{-4+(-15)}{2} = -9.5$ points per year |
| c. | *Fully conditional: principal stratification* $f(Y_i(z)\|S_i(0) > 9, S_i(1) > 9)$<br>What is the causal effect of gender on expected 3MSE scores at the fourth or ninth survey among those who would live to the ninth survey regardless of gender? | (causal effect not estimated for hypothetical data) | |
| d. | *Fully conditional: terminal decline* $f(Y_i\|S_i = s)$<br>What is the expected 3MSE two years before death? | (not estimated directly) | $-9.5$ points per year (same as row b decedents—changing time scale does not make a difference with only one stratum of decedents) |
| e. | *Partly conditional: regression conditioning on being alive* $f(Y_i\|S_i > t)$<br>What is the expected 3MSE at age 70 for people who live to be at least 70, or at age 75 for people who live to be at least 75? | $76.2 + 0.92^*5 = 80.8$ points | 0.92 points per year |
| f. | *Joint model* $f(Y_i, S_i)$<br>What is the probability of being alive and healthy at age 75 for people who were alive at age 70? | $\frac{1}{4}$ alive and 3MSE $\geq 80$ at age 75 | on average $\frac{1}{10}$ lose health/life each year (since $\frac{3}{4}$ healthy at age 70, and $\frac{1}{4}$ healthy at age 75) |

Figure 1 shows fitted 3MSE values for a baseline age of 70 years. Clearly the estimators (functions of sample data) are different for the 6 methods shown for finding 3MSE fitted values and slopes. However, we emphasize that the *estimands* are also different: the linear slope in an unconditional model is not the same as the linear slope in a fully conditional model. We assume each estimator is unbiased for its estimand, and focus on interpretation of the estimands.

### 4.1 $f(Y_i)$ Unconditional

An unconditional model, $f(Y_i)$, is appropriate if deaths do not occur, are independent of the response process, or do not result in truncation (if the response has a well-defined value following death). If these stipulations are not met, the unconditional distribution $f(Y_i)$ reflects averaging $f(Y_i|S_i)$ over the survival function $f(S_i)$, as demonstrated below. Unconditional regression models cognitive functioning at all timepoints as if nobody died, in an "immortal cohort" (Dufouil, Brayne and Clayton, 2004). The unconditional average 3MSE at age 75 years in the CHS hypothetical data is as follows:

$$\begin{aligned}
&\mathrm{E}(Y|\text{age}=75)\\
&= \mathrm{E}(Y|\text{age}=75, S>75) \cdot \mathrm{P}(S>75)\\
&\quad + \mathrm{E}(Y|\text{age}=75, S\leq 75) \cdot \mathrm{P}(S\leq 75)\\
&= \frac{90+74}{2} \cdot \frac{2}{4} + ? \cdot \frac{2}{4}.
\end{aligned}$$

For analysis methods for which "missing at random" nonresponse mechanisms are ignorable—such as for likelihood-based models fit to unbalanced longitudinal data—the value of ? is imputed implicitly because of the structure imposed by correlation between each subject's longitudinal observations (Laird, 1988). We convey this implicit imputation by projecting the hypothetical 3MSE data of deceased participants based on individual slopes. This extrapolation is more extreme than estimates would be with real data, since estimation of fixed effects in random effects models will be influenced by regression to the mean and other shrinkage (Robinson, 1991). Extending the Table 1 trajectories for Participants C and D linearly, the "incomplete" response vectors $(84, 80, 76)$ and $(65, 50, 35)$ are imputed to $(84, 80, 76, 72, 68, 64)$ and $(65, 50, 35, 20, 5, -10)$. Par-ticipant D's imputed response at age 75 ($-10$ points) is inappropriate, outside the range of the 3MSE.

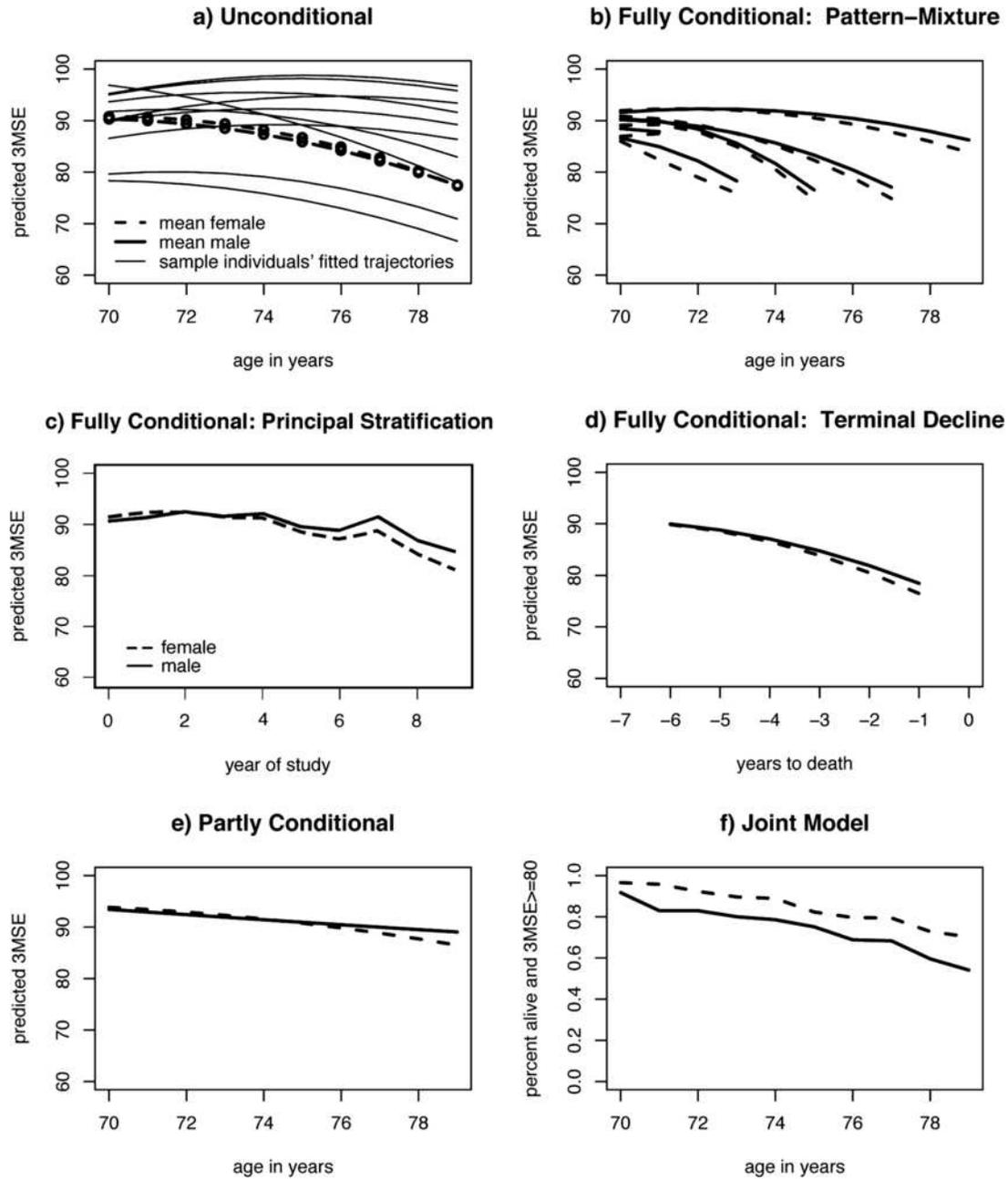

Fig. 1. *Fitted Modified Mini-Mental State Examination (3MSE) trajectories for CHS participants aged 70 years at baseline, using different models to summarize longitudinal response (3MSE) and survival. Fitted values over time are shown as solid lines for males, and as dashed lines for females.*

An unconditional model uses both observed and implicitly imputed data to estimate linear 3MSE slope and 3MSE at age 75 (Table 3, row a). Completing the estimate of 3MSE at age 75,

$$\begin{aligned}\text{E}(Y|\text{age}=75) &= \frac{90+74}{2}\cdot\frac{2}{4}+\frac{64+(-10)}{2}\cdot\frac{2}{4}\\&=54.5.\end{aligned}$$

The age 75 fitted 3MSE (54.5 points) and linear 3MSE slope (5.25 point decline per year) both yield lower estimates of cognitive functioning at age 75 than are observed for participants alive at age 75, because 3MSE values are imputed beyond death.

The unconditional model of 3MSE scores by age in the CHS data (Figure 1a) is a random effects linear regression, separating the effects of age and aging (Neuhaus and Kalbfleisch, 1998) in a quadratic model with random intercept and first-order polynomial:

$$\begin{aligned}3\text{MSE}_{ij} &= \beta_0 + b_{0i} + \beta_1\cdot\text{male}_i + \beta_2\cdot\text{age0}_i \\&\quad+ (\beta_3 + b_{1i})\cdot\text{year}_{ij} + \beta_4\cdot\text{year}_{ij}^2 \\&\quad+ \beta_5\cdot\text{male}_i\cdot\text{year}_{ij} + \beta_6\cdot\text{male}_i\cdot\text{year}_{ij}^2 \\&\quad+ \varepsilon_{ij},\end{aligned}$$

where $\text{age0}_i$ is the baseline age of participant $i$ (in years), and $\text{year}_{ij}$ is the study year (years since baseline age) for participant $i$ at timepoint $j$. $\text{male}_i$ is a dichotomous variable (with value 0 if the participant is female and value 1 if the participant is male), and random intercept, slope and error ($b_{0i}$, $b_{1i}$, and $\varepsilon_{ij}$) are normally distributed with mean 0. Likelihood-based methods such as random effects regression will fit an unconditional model, since they treat any imbalance in the data as "missing at random." Interactions between sex and linear and quadratic terms are included as potential predictors of interest. Figure 1a shows the fitted "average" trajectories for males and females (thick lines, $b_{0i} = b_{1i} = \varepsilon_{ij} = 0$), and several fitted trajectories for individuals (thin lines, selected $b_{0i}$ and $b_{1i}$). The random intercepts and slopes allow a wide range of individual fitted trajectories. However, time trends and other covariate effects are generally interpreted based on the mean model (thick lines), which reflects both observed data and data implicitly imputed beyond death. According to this average trajectory, the expected 3MSE for both males and females is 86 points at age 75, and 77 points at age 79. The unconditional model suggests rather strong age-associated declines in cognitive functioning, but the estimand is the trajectory for an immortal cohort in which truncation by death does not occur.

In rare cases, implicit imputation beyond death may be reasonable. When evaluating local recurrence following ablation of liver tumors, some livers may "die" due to transplant. Tumor marker levels ($Y_i$) and transplant candidacy ($S_i$) are related. However, since the transplant rate will never be 100%, an unconditional model and implicit imputation beyond transplant may be valid. The research question addressed is, "What would the tumor marker levels be if no transplants had occurred, but the markers had continued along the exact path that led to transplant?" While this question is relevant to liver transplants, the CHS examples show that unconditional models are generally inappropriate for longitudinal data with considerable imbalance due to death.

### 4.2 $f(Y_i|S_i = s)$ Fully Conditional

Sometimes only subjects who survive to the end of the study are included in analysis, or decedents are analyzed separately from nondecedents. An analogy in the missing data literature is pattern-mixture models (Little, 1995; Fitzmaurice and Laird, 2000), which stratify by the time of dropout. Pattern-mixture models may be fit using the same methods as unconditional models (random effects regression, etc.) but are made fully conditional by fitting separate regression models to strata defined by time of death. Generally a categorical variable defined by survival time is used as a main effect (and interaction term) in regression models, so that longitudinal trajectories are fit for groups defined by time of death (Ribaudo, Thompson and Allen-Mersh, 2000; Pauler, McCoy and Moinpour, 2003). An advantage of this approach is accurate representation of individuals' scores over time. Principal stratification (Frangakis and Rubin, 2002; Rubin, 2006) examines causal effects by using counterfactual survival to create strata. Another stratification based on time of death examines the "dying process" for decedents only, with years until death as the timescale for examining terminal decline (Siegler, 1975; Diehr et al., 2002; Wilson et al., 2003).

4.2.1 *Pattern-mixture.* Computing linear slopes separately for decedents and survivors in the hypothetical CHS data (Table 3, row b) demonstrates minimal decline in survivors [1 point per year] and terminal decline for decedents [row b(2), 9.5 points per year]. However, since the time of death is not known in advance, these models could not be used to predict an individual's trajectory based on baseline information.

In a pattern-mixture model fit to CHS data, quadratic time trends are stratified by year of death relative to baseline. Figure 1b shows fitted mean 3MSE trajectories for a baseline age of 70 years. The pattern-mixture model demonstrates terminal decline in participants who died (fitted lines that end before age 79), and reasonably stable cognitive functioning in participants who survived. The fitted 3MSE at age 75 ranged from 75 points in males who died by age 76, to 91 points in males and females who enrolled at age 70 and survived at least to age 79. The fitted mean trajectories are closer to trajectories observed for individuals than the unconditional model, but require conditioning on survival time, which is not known at baseline.

4.2.2 *Principal stratification.* Another fully conditional approach is to estimate causal effects for selected principal strata defined by potential survival outcomes (Frangakis and Rubin, 2002; Hayden, Pauler and Schoenfeld, 2005; Egleston et al., 2007; Egleston, Scharfstein and MacKenzie, 2009). Men are generally more likely to die than women at any given age. Do women have lower incidence of mental and physical decline, or are they able to survive with greater deficits than men? By examining only a hypothetical group that would survive to a certain timepoint regardless of gender, we can decouple the association of gender and cognitive function from the association of gender and survival. We estimate the association of gender and cognitive function only in the stratum of patients expected to live nine years from age 70–75 regardless of gender. The interpretation of these fully conditional models differs from pattern-mixture models, because principal stratification models are conditioned on both observed and counterfactual survival status. Generally, untestable assumptions are necessary to identify effects within the principal strata.

Specification of principal stratification models requires additional notation to describe potential outcomes. For $z = \{0$ for women, $1$ for men$\}$, let $D_i(z)$ be an indicator for potential death by a specified time and $Y_i(z)$ denote potential 3MSE scores. For this analysis, $D_i(z) = 1$ if a person dies within 9 years of study enrollment, and 0 otherwise. Let $S_i(z)$ similarly denote the potential survival time for person $i$ with gender $z$, and $S_i(z)$ determines the dimension of $Y_i(z)$. We are interested in the 3MSE scores among those in whom $D_i(0) = 0$ and $D_i(1) = 0$. Let $X_i$ represent a vector of potentially confounding covariates; we only include age as a confounding covariate. To identify potential outcomes at each time point, we make the explainable nonrandom survival assumption and use the estimator of Hayden, Pauler and Schoenfeld (2005). Explainable nonrandom survival assumes that $D_i(z) \perp D_i(1-z)|X_i$ and $D_i(z) \perp Y_i(1-z)|X_i, D_i(1-z) = 0$ where $\perp$ represents conditional independence. This assumption heuristically states that counterfactual outcomes will be conditionally independent of the observed outcomes. We need such an assumption since we do not observe counterfactual outcomes for study participants; for example, we do not observe 3MSE scores that women in the study would have had if they had been born men. We also make a pseudo-randomization assumption that the potential outcomes are independent of gender given $X_i$ as detailed in Egleston et al. (2007). Pseudo-randomization basically states that the "assignment" of gender is analogous to a random coin toss given a set of confounding covariates. We appreciate that some might feel that gender is inappropriate for such a causal analysis even as a didactic example, since gender cannot be manipulated (Holland, 1986).

Figure 1c shows fitted 3MSE scores for participants aged 70–75 at baseline who would survive 9 years regardless of gender. Men and women are both predicted to have stable scores on average for four years after baseline, and to decline an average of 3.7 points in men and 5.0 points in women over the following five-year period. Trajectories for men and for women look similar to the survivor cohort in the pattern-mixture model (Figure 1b). In other words, when survival is held comparable for both genders, men may show an advantage in cognitive functioning. This could reflect an ability for women to survive when cognitive functioning is diminished. Cognitive impairment may occur at the same rate in men and women, but may be associated with greater mortality risk in men.

One concern with this approach is that explainable nonrandom survival is a very strong assumption. A number of sensitivity analysis approaches are available to investigate whether deviations from this assumption could influence inferences (Hayden, Pauler and Schoenfeld, 2005; Egleston et al., 2007; Egleston, Scharfstein and MacKenzie, 2009). Still, some investigators have advocated that the use of relatively strong assumptions to uniquely identify principal strata effects is justifiable (Joffe, Small and Hsu, 2007; Elliott, Joffe and Chen, 2006).

The principal stratification approach is better suited to estimating causal effects for interventions than to describing the prognosis for individuals. Survival status and nonidentifiable assumptions about counterfactual survival are both part of the estimand for causal effects within each principal stratum.

4.2.3 *Terminal decline.* A third fully conditional CHS model examines terminal decline. Rather than counting forward in years of age, the time scale for this analysis counts backward from death. The 2458 participants who are alive at the end of follow-up (64%) are excluded, since their age at death is not known. As in earlier models, a random effects model is fit with quadratic time and random intercept and slope. However, the time scale is changed:

$$\begin{aligned} 3\text{MSE}_{ij} &= \beta_0 + b_{0i} + \beta_1 \cdot \text{male}_i + \beta_2 \cdot \text{age0}_i \\ &\quad + (\beta_3 + b_{1i}) \cdot \text{yr\_fr\_death}_{ij} \\ &\quad + \beta_4 \cdot \text{yr\_fr\_death}_{ij}^2 \\ &\quad + \beta_5 \cdot \text{male}_i \cdot \text{yr\_fr\_death}_{ij} \\ &\quad + \beta_6 \cdot \text{male}_i \cdot \text{yr\_fr\_death}_{ij}^2 + \varepsilon_{ij}, \end{aligned}$$

where yr_fr_death ranges from $-1$ (one year before the year of death) to $-9$ (9 years before), and other variables are as described in Section 4.1. Figure 1d shows fitted average terminal decline trajectories for men and women with baseline age 70. The estimated rate of decline is about 4.7 points per year, plus the effect of a negative quadratic coefficient ($-0.3$). The fitted 3MSE score 6 years before death is about 87 points, close to the average baseline value (88 points) for the 70-year-olds at baseline. For this cohort of decedents, averaging over yr_fr_death and male, the expected 3MSE at age 75 is 85 points, and at age 79 is 82 points. The rate of terminal decline reflects the combined influence of the most dramatic declines and the more stable 3MSE patterns observed in the multiple pattern-mixture fitted trajectories for decedents.

### 4.3 $f(Y_i|S_i > t)$ Partly Conditional

For partly conditional models, the expected value of $Y_{ij}$ (response of subject $i$ at time $t_{ij}$) conditions on the subjects being alive at time $t_{ij}$. This conditioning may seem trivial: after all, data are not collected posthumously. However, as is well documented for data missing due to nonresponse (Laird, 1988), analysis methods that model the correlation structure of longitudinal data (such as mixed models) will implicitly impute responses, whether missing due to dropout or death (Dufouil, Brayne and Clayton, 2004; Kurland and Heagerty, 2005). Partly conditional regression models may be estimated by assuming independence among the longitudinal responses. In that sense, they may be fit using linear regression or generalized linear models. However, generalized estimating equations (with independence working correlation) allow estimation of sandwich standard errors.

Partly conditional "regression conditioning on being alive" (Kurland and Heagerty, 2005) (RCA) describes 3MSE scores at different ages among the surviving participants. For the hypothetical CHS data, RCA estimates of average 3MSE at age 75 and linear 3MSE slope are calculated using linear regression (Table 3, row e). Implicit imputation is avoided by treating observations from the same person as independent. RCA accurately shows that the prevalent cognitive functioning level is slightly higher at age 75 (82 points is the average 3MSE for survivors A and B, estimated as 80.8 by imposing a single linear slope to all observed data), compared to age 70 (80.8 point average for A–D, estimated as 76.2 by linear regression). The partly conditional slope predicts that average 3MSE increases 0.92 points per year, despite that no individuals have increasing 3MSE scores. Partly conditional regression reflects 3MSE in the dynamic cohort of survivors, not individual subjects change in cognitive functioning.

The partly conditional RCA model avoids implicit imputation of data for deceased subjects, and describes longitudinal 3MSE for the dynamic cohort of survivors (Dufouil, Brayne and Clayton, 2004; Kurland and Heagerty, 2005). The regression equation is similar to that for the unconditional and pattern-mixture models, but does not include a random intercept or slope:

$$\begin{aligned} 3\text{MSE}_{ij} &= \beta_0 + \beta_1 \cdot \text{male}_i + \beta_2 \cdot \text{age0}_i + \beta_3 \cdot \text{year}_{ij} \\ &\quad + \beta_4 \cdot \text{year}_{ij}^2 + \beta_5 \cdot \text{male}_i \cdot \text{year}_{ij} \\ &\quad + \beta_6 \cdot \text{male}_i \cdot \text{year}_{ij}^2 + \varepsilon_{ij}. \end{aligned}$$

Figure 1e shows expected 3MSE scores for participants who entered the study at age 70, given that they were alive at the time 3MSE was measured. The difference in expected 3MSE at different ages is smaller than for the unconditional or fully conditional models. The expected 3MSE score of participants who entered CHS at age 70 is 91 points for surviving 75-year-olds, and 87 points for surviving 79-year-olds. However, this does not imply an average decline of 1 3MSE point per year, since the average for survivors is not the same as the trajectory from following individuals. The partly conditional model tracks the prevalent average 3MSE score in the survivors at each timepoint.

### 4.4 $f(Y_i, S_i)$ Joint Model

A joint response encompassing both survival and the longitudinal response also may be of interest. A patient facing a diagnosis may ask not only the chance of 5-year survival but the chance of 5-year survival with acceptable quality of life. A joint model of the probability of being alive and healthy (Diehr et al., 1995) characterizes the status of the entire cohort with respect to a longitudinal response and death. A related approach rescales response measures to predict the probability of being alive and healthy in a prescribed amount of time, such as one year (Diehr et al., 2001a). Because the joint response (alive and healthy) is defined at all timepoints for all individuals, longitudinal data are balanced. Therefore, analysis methods (random effects, GEE, etc.) will not be affected by differential survival. Joint models also may assess treatment effects simultaneously for longitudinal response and survival (Gray and Brookmeyer, 2000; Ratcliffe, Guo and Ten Have, 2004), or integrate morbidity and mortality in utility measures such as quality-adjusted life years.

Defining 3MSE scores $\geq 80$ points as healthy, the probability of being alive and healthy at age 75 in the hypothetical CHS data is 1/4 (Table 3, row f), which reflects a decline in the cohort, since 3/4 were alive and healthy at baseline (see Table 1). Assuming a linear trend, the decline from 3/4 healthy to 1/4 healthy over 5 years reflects a rate of decline of 1/10 of the cohort losing health or life each year.

Figure 1f shows the proportion of CHS participants who are alive and healthy at ages 70–79. The percent alive and healthy (PAH) shown is a proportion reflecting the status of study participants enrolled at age 70. Modeling of PAH also is possible, as well as constructing confidence intervals around the PAH estimate (Johnson, 2002). The percent alive and healthy is not always decreasing: individuals may regain "health," such as when a low 3MSE score was due to short-term side effects of medication. Like the partly conditional model, the joint model describes the cohort, rather than trends for individuals. Unlike the partly conditional model, the entire cohort is described at each timepoint, not only survivors. (The end of follow-up for the minority recruitment group could affect differences seen between ages 70–76 and 77–79. We avoid implicit imputation beyond 6 years by computing the empirical PAH as a simple proportion.)

For participants aged 70 years at baseline, the probability of being alive and having 3MSE $\geq 80$ at age 75 is 0.82 for females and 0.75 for males. At age 79, the PAH is 0.70 for females and 0.54 for males. Summing the area under the curve, the average years of healthy life (of 9 possible) is 7.6 for females, and 6.7 for males (Diehr et al., 1998). The average gender difference appears to be greater for the joint model than for the other fitted models in Figure 1. This reflects a survival advantage for females, which was not apparent in models that focused on the 3MSE.

### 4.5 Other Factorizations

A factorization of the joint distribution $f(Y_i, S_i)$ not yet discussed is $f(S_i|Y_i)f(Y_i)$. This framework is especially applicable to predicting survival $(S_i)$ using information from longitudinal biomarkers $(Y_i)$ (De Gruttola and Tu, 1994; Wulfsohn and Tsiatis, 1997; Ye, Lin and Taylor, 2008). Applying this model to hypothetical CHS data, we could conclude that 33% (1/3) of participants with declining 3MSE survive to age 75, while 100% (1/1) with stable 3MSE survive to age 75. This class of models would be categorized as *unconditional* in the framework discussed here, but is not considered in detail because survival, not longitudinal data, is the primary response of interest.

## 5. DISCUSSION

Through analysis of hypothetical and actual data sets, we have shown that choice of analysis has a great influence on interpretation of longitudinal data truncated by death. No single approach is appropriate in all situations, so the analysis should be chosen to address the aims of a research project.

Summaries of individual trajectories and descriptions of terminal decline are achieved with fully conditional models, in which analysis of longitudinal response is stratified by time of death. The principal stratification approach is most suited for estimating meaningful exposure or treatment effects. For example, in an investigation of the effect of trauma centers on functional outcomes (Egleston, Scharfstein and MacKenzie, 2009), principal stratification could adjust for healthy survivor bias in nontrauma centers. The partly conditional model can address situations where prevalence, rather than individual trajectories, are of interest. For example, survival information could estimate the number of new Medicare recipients who will be alive in 10 years, and a partly conditional model could then estimate the need for dementia services in those survivors. For studies of palliative care, treatment effects may be described by a joint model for longitudinal response and survival. The area under the joint density curve would summarize treatment differences in both survival and quality of life response.

Once the statistical model is clarified by research aims, choice of analysis method should be apparent (Table 2). An unconditional model is fit by random effects and other multilevel approaches, when the time scale does not depend on survival times. A fully conditional model may be fit by random effects or other analysis methods, with time scale or stratification depending on survival time (Diehr et al., 2002; Pauler, McCoy and Moinpour, 2003). Partly conditional models may be fit directly by GEE with independence working correlation (Kurland and Heagerty, 2005). Joint models may be fit as a multivariate joint distribution (Gray and Brookmeyer, 2000), or as a composite response incorporating both survival and longitudinal response (Diehr et al., 2001a).

We have focused on research in which estimation of a parameter, such as a treatment effect, slope or patient trajectory, is of primary interest. Other classes of statistical analysis are available to address different research questions. Additionally, models fit using one factorization of the joint distribution of survival and longitudinal response may be transformed to address aspects of another model. For example, a fully conditional model may be marginalized (Heagerty and Zeger (2000)) to estimate a partly conditional estimand, such as the expected 3MSE among CHS survivors at age 85 (Fitzmaurice and Laird, 2000); a joint model may estimate fully conditional trajectories (Ribaudo, Thompson and Allen-Mersh, 2000).

We have not addressed situations involving interval censoring or unknown survival times. While a two-stage imputation method (Harel et al. (2007)) could address both nonresponse and missing survival information, the data analysis method should be chosen carefully to avoid implicit imputation and to address research aims.

In longitudinal studies in which some subjects die yet another response, such as cognitive functioning or quality of life, is of primary interest, careful modeling is required to identify an analysis method to address research aims. When deaths occur at many different times along the time frame for which responses are measured (i.e., age or time from baseline), random effects models (which are unconditional with respect to survival) may implicitly impute data beyond death. Implicit imputation is a fundamental strength of random effects models in the missing data context, but limits the suitability of these unconditional models in analyzing longitudinal data with great imbalance due to death. When the time scale describes time from (not until) death, the model becomes fully conditional. For terminal decline models, implicit imputation beyond death will not occur when random effects models are fit. Analysts concerned about the potential impact of implicit imputation may fit a generalized linear model or generalized estimating equations with independence correlation (which fit partly conditional models) and compare fitted parameters to an unconditional model.

## ACKNOWLEDGMENTS

This research was supported in part by the Intramural Research Program of the NIH, by NIH Grant P30 CA 06927 and by an appropriation from the Commonwealth of Pennsylvania. Research reported in this article was supported by contract numbers N01-HC-85079 through N01-HC-85086, N01-HC-35129, N01 HC-15103, N01 HC-55222, N01-HC-75150, N01-HC-45133, Grant number U01 HL080295 from the National Heart, Lung, and Blood Institute, with additional contribution from the National Institute of Neurological Disorders and Stroke. A full list of principal CHS investigators and institutions can be found at http://www.chs-nhlbi.org/pi.htm.

## REFERENCES

Burke, G. L., Arnold, A. M., Bild, D. E., Cushman, M., Fried, L. P., Newman, A., Nunn, C. and Robbins, J.

(2001). Factors associated with healthy aging: the cardiovascular health study. *Journal of the American Geriatrics Society* **49** 254–262.

DE GRUTTOLA, V. and TU, X. M. (1994). Modelling progression of CD4-lymphocyte count and its relationship to survival time. *Biometrics* **50** 1003–1014.

DIEHR, P., PATRICK, D., HEDRICK, S., ROTHMAN, M., GREMBOWSKI, D., RAGHUNATHAN, T. E. and BERESFORD, S. (1995). Including deaths when measuring health status over time. *Medical Care* **33** AS164–AS172.

DIEHR, P., PATRICK, D. L., BILD, D. E., BURKE, G. L. and WILLIAMSON, J. D. (1998). Predicting future years of healthy life for older adults. *Journal of Clinical Epidemiology* **51** 343–353.

DIEHR, P., PATRICK, D. L., SPERTUS, J., KIEFE, C. I., MCDONELL, M. and FIHN, S. D. (2001a). Transforming self-rated health and the SF-36 scales to include death and improve interpretability. *Medical Care* **39** 670–680.

DIEHR, P., WILLIAMSON, J., PATRICK, D. L., BILD, D. E. and BURKE, G. L. (2001b). Patterns of self-rated health in older adults before and after sentinel health events. *Journal of the American Geriatrics Society* **49** 36–44.

DIEHR, P., WILLIAMSON, J., BURKE, G. L. and PSATY, B. M. (2002). The aging and dying processes and the health of older adults. *Journal of Clinical Epidemiology* **55** 269–278.

DUFOUIL, C., BRAYNE, C. and CLAYTON, D. (2004). Analysis of longitudinal studies with death and drop-out: A case study. *Stat. Med.* **23** 2215–2226.

EGLESTON, B. L., SCHARFSTEIN, D. O., FREEMAN, E. E. and WEST, S. K. (2007). Causal inference for non-mortality outcomes in the presence of death. *Biostatistics* **8** 526–545.

EGLESTON, B. L., SCHARFSTEIN, D. O. and MACKENZIE, E. (2009). On estimation of the survivor average causal effect in observational studies when important confounders are missing due to death. *Biometrics* **65** 497–504.

ELLIOTT, M. R., JOFFE, M. M. and CHEN, Z. (2006). A potential outcomes approach to developmental toxicity analyses. *Biometrics* **62** 352–360. MR2227484

FITZMAURICE, G. M. and LAIRD, N. M. (2000). Generalized linear mixture models for handling nonignorable dropouts in longitudinal studies. *Biostatistics* **1** 141–156.

FRANGAKIS, C. E. and RUBIN, D. B. (2002). Principal stratification in causal inference. *Biometrics* **58** 21–29. MR1891039

FRANGAKIS, C. E., RUBIN, D. B., AN, M. W. and MACKENZIE, E. (2007). Principal stratification designs to estimate input data missing due to death. *Biometrics* **63** 641–649; discussion 650–662. MR2395697

FRIED, L. P., BORHANI, N. O., ENRIGHT, P., FURBERG, C. D., GARDIN, J. M., KRONMAL, R. A., KULLER, L. H., MANOLIO, T. A., MITTELMARK, M. B., NEWMAN, A., O'LEARY, D. H., PSATY, B., RAUTAHARJU, P., TRACY, R. P., WEILER, P. G. and RESEARCH GROUP (CHS) (1991). The cardiovascular health study: Design and rationale. *Annals of Epidemiology* **1** 263–276.

GRAY, S. M. and BROOKMEYER, R. (2000). Multidimensional longitudinal data: estimating a treatment effect from continuous, discrete, or time-to-event response variables. *J. Amer. Statist. Assoc.* **95** 396–406.

HAREL, O., HOFER, S. M., HOFFMAN, L., PEDERSEN, N. L. and JOHANSSON, B. (2007). Population inference with mortality and attrition in longitudinal studies on aging: A two-stage multiple imputation method. *Experimental Aging Research* **33** 187–203.

HAYDEN, D., PAULER, D. K. and SCHOENFELD, D. (2005). An estimator for treatment comparisons among survivors in randomized trials. *Biometrics* **61** 305–310. MR2135873

HEAGERTY, P. J. and ZEGER, S. L. (2000). Marginalized multilevel models and likelihood inference (with discussion). *Statist. Sci.* **15** 1–26. MR1842235

HOLLAND, P. W. (1986). Statistics and causal inference (C/R: P961–P970). *J. Amer. Statist. Assoc.* **81** 945–960. MR0867618

JOFFE, M. M., SMALL, D. and HSU, C. Y. (2007). Defining and estimating intervention effects for groups that will develop an auxiliary outcome. *Statist. Sci.* **22** 74–97. MR2408662

JOHNSON, L. L. (2002). Incorporating death into the statistical analysis of categorical longitudinal health status data. Ph.D. thesis, Univ. Washington.

KAPLAN, R. C., TIRSCHWELL, D. L., LONGSTRETH, W. T., J., MANOLIO, T. A., HECKBERT, S. R., LEFKOWITZ, D., EL-SAED, A. and PSATY, B. M. (2005). Vascular events, mortality, and preventive therapy following ischemic stroke in the elderly. *Neurology* **65** 835–842.

KURLAND, B. F. and HEAGERTY, P. J. (2004). Marginalized transition models for longitudinal binary data with ignorable and non-ignorable drop-out. *Stat. Med.* **23** 2673–2695.

KURLAND, B. F. and HEAGERTY, P. J. (2005). Directly parameterized regression conditioning on being alive: Analysis of longitudinal data truncated by deaths. *Biostatistics* **6** 241–258.

LAIRD, N. M. (1988). Missing data in longitudinal studies. *Stat. Med.* **7** 305–315.

LAIRD, N. M. and WARE, J. H. (1982). Random-effects models for longitudinal data. *Biometrics* **38** 963–974.

LIANG, K. Y. and ZEGER, S. L. (1986). Longitudinal data analysis using generalized linear models. *Biometrika* **7** 13–22. MR0836430

LITTLE, R. J. A. (1995). Modeling the drop-out mechanism in repeated-measures studies. *J. Amer. Statist. Assoc.* **90** 1112–1121. MR1354029

LITTLE, R. J. A. and RUBIN, D. B. (1987). *Statistical Analysis with Missing Data*. Wiley, New York. MR0890519

NEUHAUS, J. M. and KALBFLEISCH, J. D. (1998). Between- and within-cluster covariate effects in the analysis of clustered data. *Biometrics* **54** 638–645.

PAIK, M. C. (1997). The generalized estimating equation approach when data are not missing completely at random. *J. Amer. Statist. Assoc.* **92** 1320–1329.

PAULER, D. K., MCCOY, S. and MOINPOUR, C. (2003). Pattern mixture models for longitudinal quality of life studies in advanced stage disease. *Stat. Med.* **22** 795–809.

RATCLIFFE, S. J., GUO, W. and TEN HAVE, T. R. (2004). Joint modeling of longitudinal and survival data via a common frailty. *Biometrics* **60** 892–899. MR2133541

Ribaudo, H. J., Thompson, S. G. and Allen-Mersh, T. G. (2000). A joint analysis of quality of life and survival using a random effect selection model. *Stat. Med.* **19** 3237–3250.

Robins, J. M., Rotnitzky, A. and Zhao, L. P. (1995). Analysis of semiparametric regression models for repeated outcomes in the presence of missing data. *J. Amer. Statist. Assoc.* **90** 106–121. MR1325118

Robinson, G. K. (1991). That BLUP is a good thing: The estimation of random effects (with discussion). *Statist. Sci.* **6** 15–51. MR1108815

Rubin, D. B. (1987). *Multiple Imputation for Nonresponse in Surveys*. Wiley, New York. MR0899519

Rubin, D. B. (2006). Causal inference through potential outcomes and principal stratification: Application to studies with "censoring" due to death. *Statist. Sci.* **21** 299–309. MR2339125

Siegler, I. C. (1975). The terminal drop hypothesis: Fact or artifact? *Experimental Aging Research* **1** 169–185.

Teng, E. L. and Chui, H. C. (1987). The Modified Mini-Mental State (3MS) examination. *Journal of Clinical Psychiatry* **48** 314–318.

Wilson, R. S., Beckett, L. A., Bienias, J. L., Evans, D. A. and Bennett, D. A. (2003). Terminal decline in cognitive function. *Neurology* **60** 1782–1787.

Wulfsohn, M. S. and Tsiatis, A. A. (1997). A joint model for survival and longitudinal data measured with error. *Biometrics* **53** 330–339. MR1450186

Ye, W., Lin, X. and Taylor, J. M. G. (2008). Semiparametric modeling of longitudinal measurements and time-to-event data—a two-stage regression calibration approach. *Biometrics* **64** 1238–1246.